\documentclass[prb,aps,twocolumn,preprintnumbers,amsmath,amssymb,superscriptaddress,altaffilletter]{revtex4}

\usepackage[utf8]{inputenc}
\usepackage{array}
\usepackage{amsmath}
\usepackage{multirow}
\usepackage{graphicx}
\usepackage{dcolumn}

\begin{document}


\title[On-chip Dual-Comb based on Quantum Cascade Lasers]{On-chip Dual-comb based on Quantum Cascade Laser Frequency Combs}

\author{G. Villares}
\email{gustavo.villares@phys.ethz.ch}
\affiliation{ 
	Institute for Quantum Electronics, ETH Z\"urich, CH-8093 Z\"urich, Switzerland
}%
\author{J. Wolf}%
\affiliation{ 
	Institute for Quantum Electronics, ETH Z\"urich, CH-8093 Z\"urich, Switzerland
}%
\author{D. Kazakov}%
\affiliation{ 
	Institute for Quantum Electronics, ETH Z\"urich, CH-8093 Z\"urich, Switzerland
}%
\author{M. J. S\"uess}%
\affiliation{ 
	Institute for Quantum Electronics, ETH Z\"urich, CH-8093 Z\"urich, Switzerland
}%
\author{A. Hugi}%
\affiliation{ 
	IRsweep GmbH, CH-8093 Z\"urich, Switzerland
}%
\author{M. Beck}%
\affiliation{ 
	Institute for Quantum Electronics, ETH Z\"urich, CH-8093 Z\"urich, Switzerland
}%
\author{J. Faist}
\email{jfaist@phys.ethz.ch}
\affiliation{ 
	Institute for Quantum Electronics, ETH Z\"urich, CH-8093 Z\"urich, Switzerland
}%

\date{\today}

\begin{abstract}
Dual-comb spectroscopy is emerging as one of the most appealing applications of mid-infrared frequency combs for high-resolution molecular spectroscopy, as it leverages on the unique coherence properties of frequency combs combined with the high sensitivities achievable by mid-infrared molecular spectroscopy.
Here we present an on-chip dual-comb source based on mid-infrared quantum cascade laser frequency combs, where two frequency combs are integrated on a single chip. Control of the combs repetition and offset frequencies is obtained by integrating micro-heaters next to each laser. We show that a full control of the dual-comb system is possible, by measuring a multi-heterodyne beating corresponding to an optical bandwidth of 32\,cm$^{-1}$ at a center frequency of 1330\,cm$^{-1}$ (7.52\,$\mu$m), demonstrating that this device is ideal for compact dual-comb spectroscopy systems.    
\end{abstract}

\maketitle

Optical sensing by means of frequency combs~\cite{udem_optical_2002} is seen as an attractive spectroscopy tool as it combines high accuracy and precision together with a wide spectral coverage~\cite{diddams2010evolving}. Dual-comb spectroscopy~\cite{schiller2002spectrometry,keilmann2004time,coddington2008coherent,coddington2010coherent} is becoming one of the most attractive spectroscopy techniques based on frequency combs, as all the comb spectrum can be acquired in very short time scales without requiring any moving parts.

Although first demonstrated in the mid-infrared (MIR) part of the spectrum,
dual-comb spectroscopy has been extensively developed in the near-infrared spectral region~\cite{coddington2008coherent,bernhardt2009cavity,coddington2010coherent}, as frequency combs are mature sources in this spectral region. Extending this technique to the MIR range~\cite{schliesser2012mid}, where the fundamental roto-vibrational transitions of most gas molecules are present, will allow to achieve dual-comb spectroscopy measurements with accuracies and precisions never achieved~\cite{bernhardt2010mid,baumann2011spectroscopy,giorgetta2015broadband}. 

Nonetheless, optical sources capable of generating MIR frequency combs are extremely complex~\cite{schliesser2012mid,Adler:2009,Vodopyanov:2011,Galli:2013d,cruz2015mid} and dual-comb spectroscopy on the MIR range is only possible in highly equipped laboratories~\cite{baumann2011spectroscopy,giorgetta2015broadband}. Instead, a compact dual-comb spectrometer in the MIR range could bring broadband high-precision measurements to practical applications such as trace gas or breath analysis, just to name a few.

 Quantum Cascade Lasers (QCL)~\cite{faist1994quantum} are semiconductor lasers capable of generating frequency combs in the MIR and Terahertz parts of the spectrum~\cite{hugi2012mid,Burghoff:2014,rosch2015octave,lu2015high}. As the comb formation takes place directly in the QCL active region, QCL frequency combs (QCL-combs) offer the unique possibility of a completely integrated chip-based system capable of performing broadband high-resolution spectroscopy. In the meantime, a theoretical description of the comb formation has already been developed~\cite{Khurgin:2014,Villares:2015}, dispersion compensation of QCL-combs was investigated~\cite{Burghoff:2014,villares2015dispersion} and dual-comb spectroscopy using QCL-combs has been demonstrated~\cite{Villares:2014}. In this letter, we demonstrate an on-chip dual-comb source based on MIR QCL-combs. We show that control of offset and repetition frequencies of both combs is possible by integrating micro-heaters~\cite{bismuto2015electricaltuning} close to the QCL-combs, demonstrating that this device is ideal for compact dual-comb spectroscopy systems. 


An optical frequency comb is a coherent source whose spectrum consists of a set of equally spaced modes~\cite{udem_optical_2002}. Each comb mode $f_n$ can therefore be expressed as
\begin{equation}
	f_n = f_{\textrm{ceo}} + n f_{\textrm{rep}}
\end{equation}
 where $f_{\textrm{rep}}$ is the repetition frequency and corresponds to the spacing between the modes and $n$ is an integer number. The offset frequency $f_{\textrm{ceo}}$ corresponds to a common offset shared between all modes.
 
Dual-comb spectroscopy is based on the generation of a multi-heterodyne beating between two frequency combs with slightly different repetition frequencies ($f_{\textrm{rep},1}$ and $f_{\textrm{rep},2} = f_{\textrm{rep},1} + \Delta f_{\textrm{rep}}$, respectively, where $\Delta f_{\textrm{rep}}$ is the difference in repetition frequencies) on a fast detector. One comb is used as a local oscillator while the other is used to interrogate a sample. Each multi-heterodyne beat contains information regarding the sample absorption at the optical frequency of the comb line interrogating the sample. Therefore, the full control of both offset and repetition frequencies is necessary when designing a device for dual-comb spectroscopy applications.

\begin{figure}[!t]
	\includegraphics[width=3.37 in]{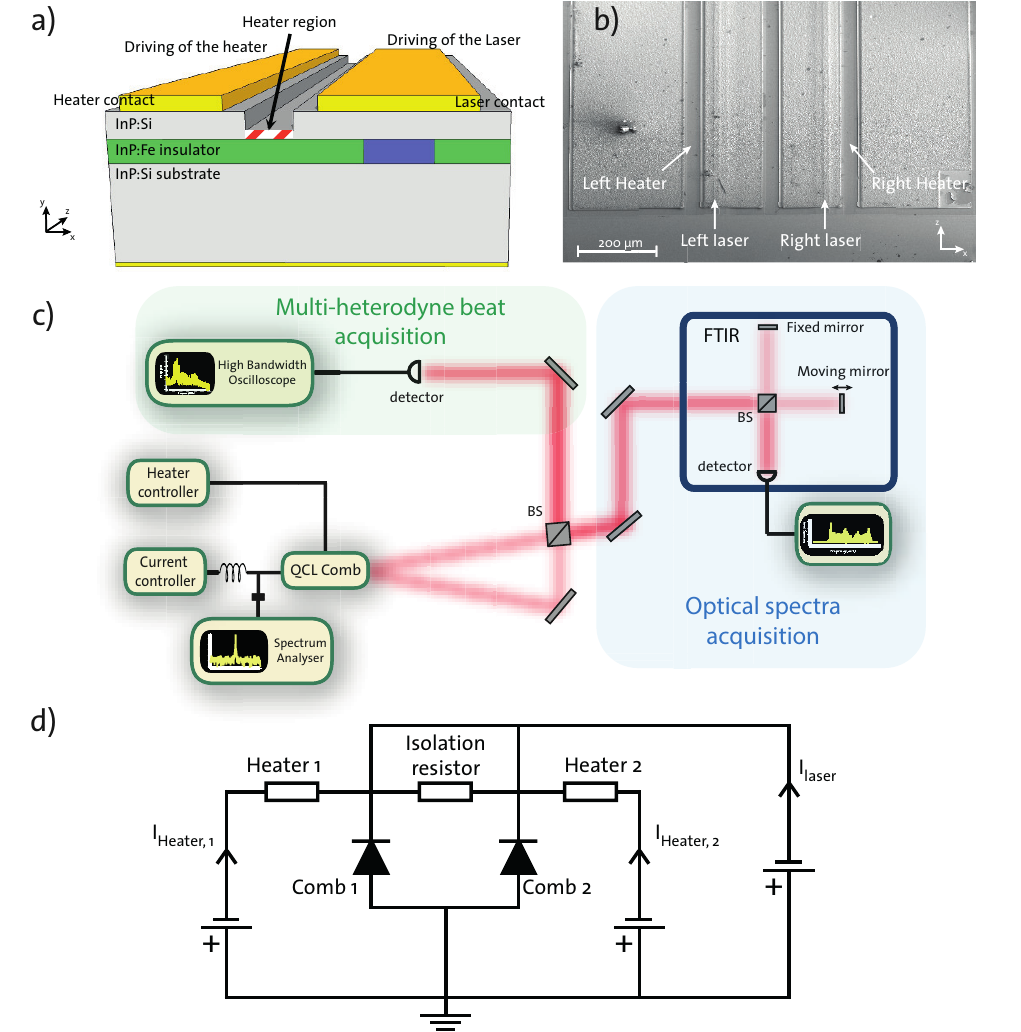}
	\caption{\label{fig:fig1} On-chip dual-comb based on QCL-combs.
		\textbf{a} Schematic representation of the chip containing a QCL-comb and a micro-heater.
		\textbf{b} Scanning electron micrograph showing the front view of the device containing two QCL-combs together with two micro-heaters.  
		\textbf{c} Optical characterization setup used for characterization of the comb optical spectra, the multi-heterodyne beating between the two combs as well as the characterization of the RF beatnote. BS: beam-splitter.
		\textbf{d} Electrical circuit of an on-chip dual-comb. The isolation resistor can be made with high values ($\sim$ M$\Omega$) in order to isolate the combs.
		}
\end{figure}

Our concept consists of two QCL-combs fabricated on a single chip, into which two independent micro-heaters~\cite{bismuto2015electricaltuning} are integrated next to each QCL-comb. A schematic representation of a QCL-comb together with its micro-heater is shown in Fig.\,\ref{fig:fig1}\textbf{a}. The QCL-comb sources used in this study are based on a modified version of an InGaAs/InAlAs broadband QCL design previously reported~\cite{hugi2012mid}. The micro-heater consists of a small resistor created by etching a thin layer of Si-doped InP, previously utilized for achieving spectral tuning of single frequency QCLs~\cite{bismuto2015electricaltuning}. The QCL-comb and the micro-heater are biased by two independent current sources, as shown schematically in Fig.\,\ref{fig:fig1}\textbf{d}. The source driving the QCL is used for achieving laser action, but also for controlling the offset and repetition frequencies of the QCL-comb. In addition to this first way of controlling the comb parameters, the source driving the micro-heater controls the amount of heat generated by the micro-heater. By using the temperature tuning of the refractive index, both offset and repetition frequencies of the QCL-comb can also be controlled by the micro-heater. These two ways of controlling the comb parameters are not strictly equivalent as the laser current controls the amount of carriers in the structure, thus its gain and its refractive index. In addition to this effect, the laser current also increases the temperature of the structure and therefore induces a modification of the refractive index. Only this second effect is utilized by the micro-heater.

 Fig.\,\ref{fig:fig1}\textbf{b} shows a scanning electron microscopy image of the fabricated device containing two QCL-combs and two micro-heaters. The QCL-combs are separated by $\simeq 200$\,$\mu$m and can be electrically isolated by a deep etched section done between the two lasers.

\begin{figure}[!b]
	\includegraphics[width=3.37 in]{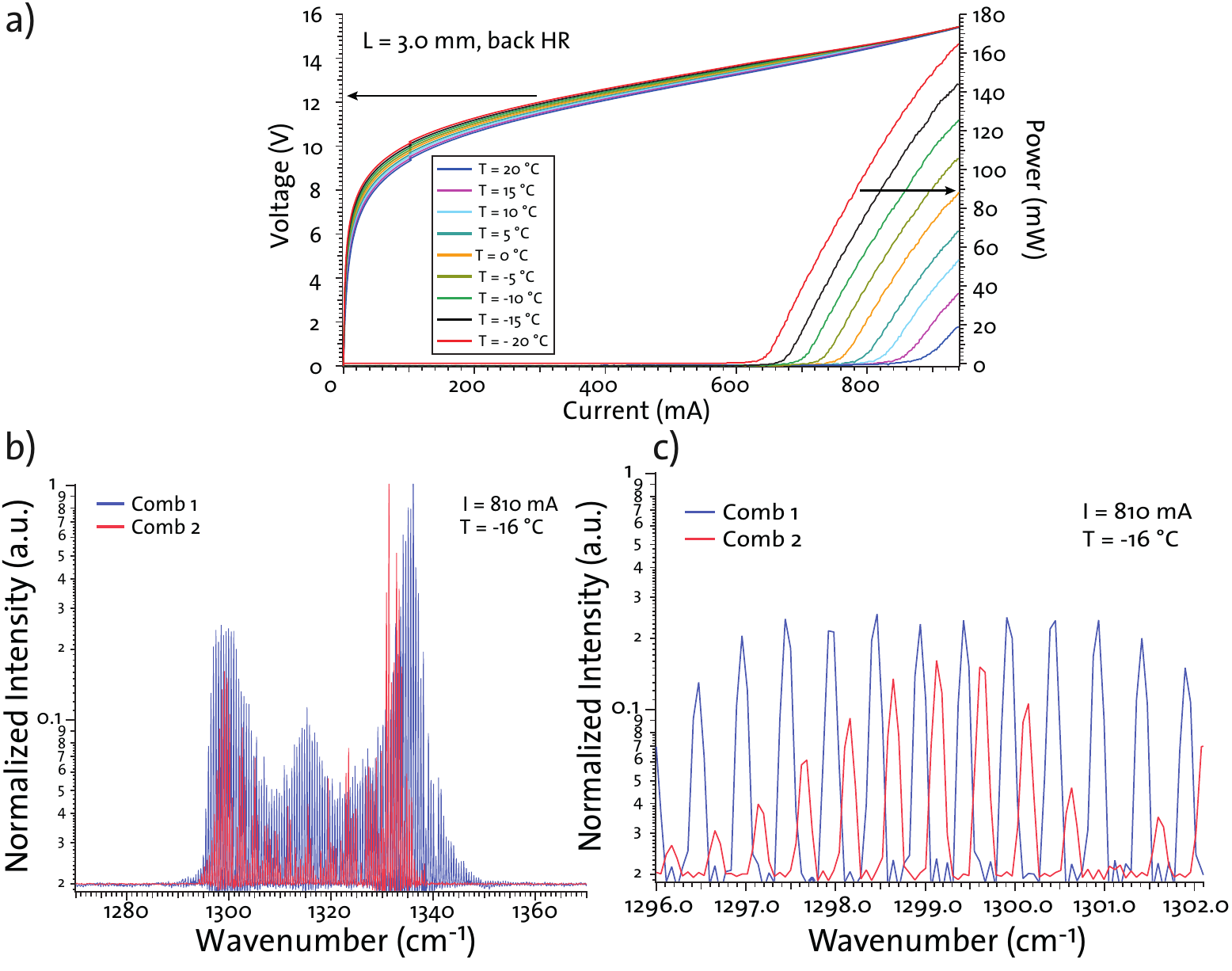}
	\caption{\label{fig:fig2} Optical characterization of an on-chip dual-comb source based on MIR QCL-combs. 
		\textbf{a} Light intensity-current-voltage characteristic of a typical device. The power represents the total power of the two combs.
		\textbf{b} Optical spectra of the two combs, acquired by using the characterization set-up represented in Fig.\,\ref{fig:fig1}\textbf{c} (acquired with a FTIR, 0.12\,cm$^{-1}$ resolution).  
		\textbf{c} Magnified view of the optical spectra represented in Fig.\,\ref{fig:fig2}\textbf{b}, showing the difference in offset frequencies between the two combs (0.12\,cm$^{-1}$ resolution).}
\end{figure}

For optical characterisation of this system, the dual-comb source is collimated by a single high-numerical aperture (0.86) aspheric lens, as shown in Fig.\,\ref{fig:fig1}\textbf{c}. Both beams are first spatially separated and then combined in a 50/50 beam-splitter, giving access to two versions of the superposition between the two beams. The first is sent to a fast detector (HgCdTe, 250\,MHz 3dB cut-off bandwidth) in order to measure the multi-heterodyne beating between the two combs. The second is sent to a Fourier Transform Infrared Spectrometer (FTIR, 0.12\,cm$^{-1}$ resolution) for acquisition of the optical spectra. The optical spectrum of each comb can be acquired independently by blocking one of the beams. For technical reasons, both lasers are driven in parallel by using a single low noise current source (Wavelength electronics QCL2000 LAB) with a specified average current noise density of 2\,nA/$\sqrt{\textrm{Hz}}$. Also, the micro-heater is driven by using a low noise current driver source (Wavelength electronics QCL1000 OEM) with the same noise characteristics.

Fig.\,\ref{fig:fig2} shows the typical performance of an on-chip dual-comb source based on MIR QCL-combs. A 3\,mm long device coated with a high-reflection coating on the back-facet operates at room-temperature emitting $>100$\,mW of output power in CW operation (c.f. Fig.\,\ref{fig:fig2}\textbf{a}). As the devices are driven in parallel, the current represents the total current circulating in both lasers. Also, the output power is the total output power of the two combs. The optical power characteristics of the individual devices were also characterized, showing similar performances for each device, with a ratio between the output power of each device $P_1/P_2$ = 0.83. The optical spectra of both devices (cf. Fig.\,\ref{fig:fig2}\textbf{b}) is centered at 1330\,cm$^{-1}$ with $\simeq$ 50\,cm$^{-1}$ of optical bandwidth. No micro-heater is used in this case. A zoom on the optical spectra shows that the offset frequencies of the two QCL-combs are significantly different when no micro-heater is used ($\simeq$ 7\,GHz in this case). The origin of this offset frequency difference $\Delta f_{\textrm{ceo}}$ lies on the precision of the lithography step used to define the laser ridges. Achieving devices with nearly identical offset frequencies ($<1$\,MHz) represents a strong requirement in terms of precision of the laser width. In addition, a high value for $\Delta f_{\textrm{ceo}}$ represents an important technological drawback for dual-comb systems, as the multi-heterodyne beat would be observed at very high frequencies ($\simeq$ 7\,GHz in this example), creating a need for high-bandwidth detectors.   

\begin{figure}[!t]
	\includegraphics[width=3.37 in]{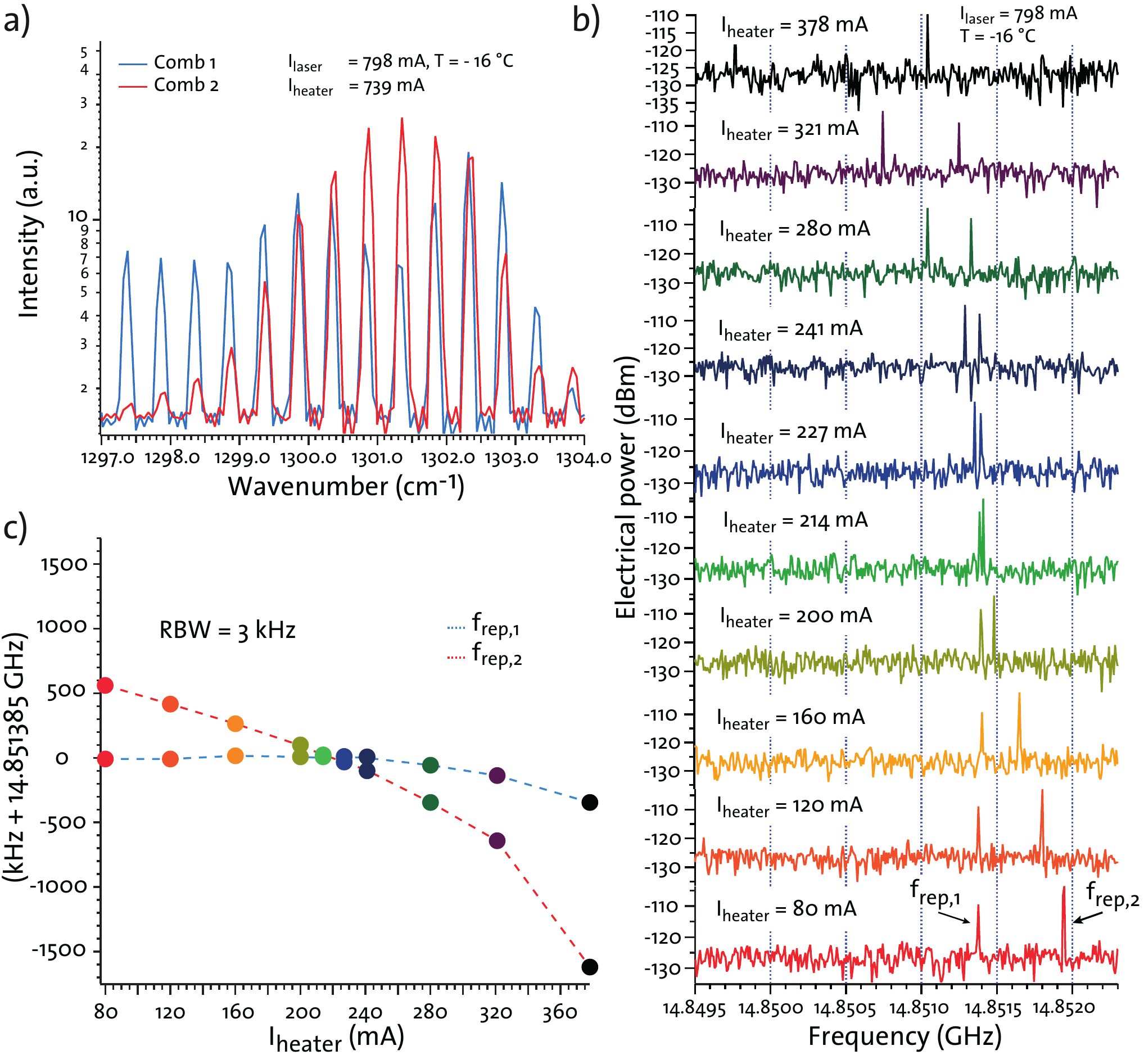}
	\caption{\label{fig:fig3} Control of the comb parameters by using the integrated micro-heaters. 
		\textbf{a} Optical spectra of the two combs when both the laser current I$_{\textrm{laser}}$ and one of the micro-heater currents I$_{\textrm{heater,1}}$ were optimized in order to minimize the $\Delta f_{\textrm{ceo}}$.
		\textbf{b} Set of RF spectra acquired for different values of I$_{\textrm{heater,1}}$, showing the two RF beatnotes corresponding to $f_{\textrm{rep,1}}$ and $f_{\textrm{rep,2}}$ (resolution bandwidth (RBW) = 3\,kHz, span = 5\,MHz, sweep time = 35\,ms.).  
		\textbf{c} Value of the frequencies $f_{\textrm{rep,1}}$ and $f_{\textrm{rep,2}}$ extracted from Fig.\,\ref{fig:fig3}\textbf{b}, showing the independent control of the repetition frequency (the dots are colored according to the spectra of Fig.\,\ref{fig:fig3}\textbf{b}).}
\end{figure}

In order to cope with this challenge, the micro-heaters were used to control the offset frequencies difference $\Delta f_{\textrm{ceo}}$ between the two QCL-combs. Fig.\,\ref{fig:fig3}\textbf{a} shows the optical spectra of the two combs when  the current driving the lasers I$_{\textrm{laser}}$ and the current driving one of the micro-heaters I$_{\textrm{heater,1}}$ are optimized to obtain almost identical offset frequencies, \emph{i.e.} $f_{\textrm{ceo,1}} \simeq f_{\textrm{ceo,2}}$. As can be observed from Fig.\,\ref{fig:fig3}\textbf{a}, the two combs seem to be aligned in frequency (within the resolution of the measurement)
and do not show a significant difference in offset frequencies as compared to the case when the heater is not used (see Fig.\,\ref{fig:fig2}\textbf{a}).  

Further control of the comb parameters is demonstrated by measuring the radio frequency (RF) spectrum of the laser current~\cite{Villares:2014}. The RF spectrum contains the beatnotes at the roundtrip frequency of the two combs $f_{\textrm{rep,1}}$ and $f_{\textrm{rep,2}}$, as they are biased in parallel by a unique current driver. One can therefore precisely measure both comb repetition frequencies with a spectrum analyzer, as shown in Fig.\,\ref{fig:fig1}\textbf{c}.

We show the control of $f_{\textrm{rep}}$ of a single comb by using one micro-heater to tune $f_{\textrm{rep,1}}$ while the second micro-heater is not utilized, therefore not acting on $f_{\textrm{rep,2}}$. The RF spectra acquired at different values of micro-heater current I$_{\textrm{heater,1}}$ (all other parameters being kept fixed) are shown in Fig.\,\ref{fig:fig3}\textbf{b}. When increasing the current in the micro-heater, we observe that one RF beatnote is shifted to lower frequencies while the other RF beatnote is not shifted, effectively decreasing $\Delta f_{\textrm{rep}}$ (observed from I$_{\textrm{heater,1}}$ = 80\,mA up to 214\,mA). Around one precise value of the micro-heater current (around 214\,mA in this example), both repetition frequencies are similar. By further increasing the value of the micro-heater current, one RF beatnote is further shifted to lower frequencies and $\Delta f_{\textrm{rep}}$ starts to increase. For important values of micro-heater current, we observe that both RF beatnotes start to shift to lower frequencies, one being more sensitive to the micro-heater current than the other. This is explained by the fact that at these values of the micro-heater current, the heat dissipated starts to influence the second comb. All these observations are summarized in Fig.\,\ref{fig:fig3}\textbf{c}, where $f_{\textrm{rep,1}}$ and $f_{\textrm{rep,2}}$ are represented as a function of I$_{\textrm{heater,1}}$. This experiment finally demonstrates the control of the repetition frequency of a single comb by employing the micro-heater. Although the control of $f_{\textrm{ceo}}$ and $f_{\textrm{rep}}$ of a single comb is possible with our system, these two quantities are yet not controllable independently.      

In order to finally demonstrate the capabilities of this system for compact dual-comb spectroscopy applications, we measure the multi-heterodyne beat signal created by the beating of the two combs on a high-bandwidth HgCdTe detector, as shown schematically in Fig.\,\ref{fig:fig1}\textbf{a}. In this case, both $\Delta f_{\textrm{ceo}}$ and $\Delta f_{\textrm{rep}}$ were optimized by controlling the laser and micro-heater currents according to the considerations detailed previously.

\begin{figure}[!t]
	\includegraphics[width=3.37 in]{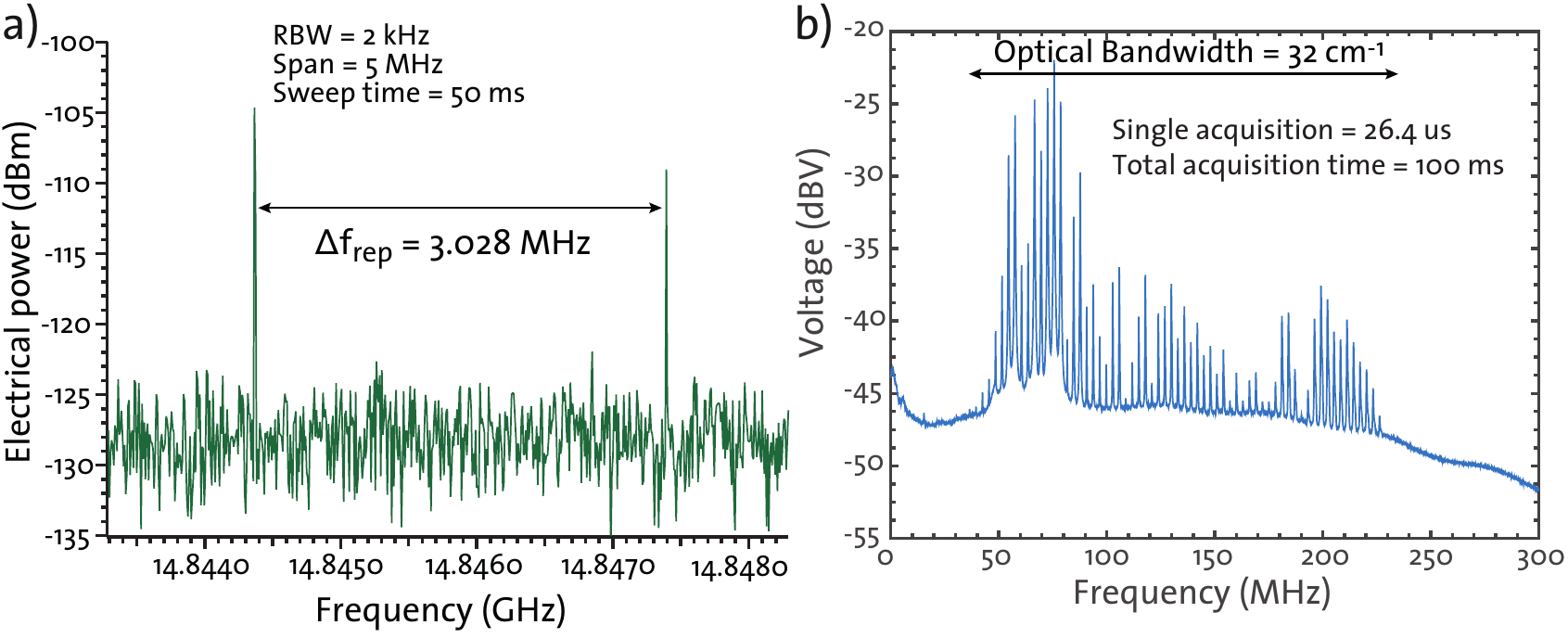}
	\caption{\label{fig:fig4} Optimization for compact dual-comb spectroscopy application. 
		\textbf{a} RF spectrum displaying the two beatnotes $f_{\textrm{rep,1}}$ and $f_{\textrm{rep,2}}$, corresponding to $\Delta f_{\textrm{rep}}=3.208$\,MHz (resolution bandwidth (RBW) = 2\,kHz, span = 5\,MHz, sweep time = 50\,ms.).
		\textbf{b} Multi-heterodyne beating of the two combs, corresponding to an optical bandwidth of 32\,cm$^{-1}$. Total acquisition time: 100\,ms. The algorithm used for averaging the RF spectra is detailed in~\cite{Villares:2014}.}
\end{figure}

Fig.\,\ref{fig:fig4} shows the optimized optical characteristics of the on-chip dual-comb based on QCL-combs. Both comb repetition frequencies were controlled in order to obtain $\Delta f_{\textrm{rep}}\simeq 3$\,MHz, as shown in Fig.\,\ref{fig:fig4}\textbf{a}. At the same time, both comb offset frequencies were optimized to be close to each other. Fig.\,\ref{fig:fig4}\textbf{b} shows the multi-heterodyne beating between the combs, showing that $\Delta f_{\textrm{ceo}}<$ 50\,MHz. This value could be further reduced by slightly adjusting the micro-heater current. Moreover, 64 modes are observed on the multi-heterodyne spectrum, corresponding to an optical bandwidth of 32\,cm$^{-1}$, as the comb repetition frequency is $\simeq$ 0.5\,cm$^{-1}$ (15\,GHz).   

A further advantage of our system is its robustness regarding temperature drifts, which is quantified by the relative frequency temperature dependence coefficient $\frac{1}{\nu}\frac{\Delta \nu_{\textrm{RF}}}{\Delta T}$ of one multi-heterodyne beatnote. In order to demonstrate the advantage of an on-chip system compared to a system where both combs would be totally independent, we measure this coefficient by using our previously demonstrated dual-comb spectrometer~\cite{Villares:2014}, where the two combs consist of two different devices. In this case, a temperature fluctuation of one comb will directly translate into a temperature drift of the multi-heterodyne beatnotes. However, in a system where both combs are on the same chip, any temperature drift of the substrate will influence both combs in the same manner, therefore minimizing the temperature drift of the multi-heterodyne beatnotes. A value of $-8.8$\,x\,$10^{-5}$\,K$^{-1}$ is measured in a system where both combs are independent, compared to a value of $-1.5$\,x\,$10^{-6}$\,K$^{-1}$ for the on-chip system, therefore demonstrating a sixty-fold increase of the robustness regarding temperature fluctuations.   

In summary, we have demonstrated an on-chip dual-comb system based on MIR QCL-combs. By employing two independent micro-heaters directly integrated next to each QCL-comb, the combs repetition and offset frequencies are controlled by utilizing the temperature tuning of the refractive index. 
As a first step for compact dual-comb spectroscopy, we demonstrate that a multi-heterodyne beating originating from the beating of the two combs can be observed and that the parameters $\Delta f_{\textrm{ceo}}$ and $\Delta f_{\textrm{rep}}$ can be precisely controlled. We show a multi-heterodyne beating corresponding to an optical bandwidth of 32\,cm$^{-1}$ with a $\Delta f_{\textrm{rep}}$ of 3\,MHz, centered at 1330\,cm$^{-1}$. Finally, due to the proximity of the micro-heaters to the QCL active region, these devices can achieve kHz of modulation bandwidth~\cite{bismuto2015electricaltuning}. This system can therefore be utilized for achieving a fully stabilized compact dual-comb spectrometer by employing, for example, all-electrical frequency noise stabilization techniques~\cite{sergachev2014all}. 

\begin{acknowledgments}
We thank Dr. Pierre Jouy for fruitful discussions. This work was financially supported by the Swiss National Science Foundation (SNF200020 - 152962), by the ETH Pioneer Fellowship programme as well as by the DARPA program SCOUT (W31P4Q-15-C-0083).
\end{acknowledgments}

\bibliography{MyLibrary}

\begin{thebibliography}{26}
\expandafter\ifx\csname natexlab\endcsname\relax\def\natexlab#1{#1}\fi
\expandafter\ifx\csname bibnamefont\endcsname\relax
  \def\bibnamefont#1{#1}\fi
\expandafter\ifx\csname bibfnamefont\endcsname\relax
  \def\bibfnamefont#1{#1}\fi
\expandafter\ifx\csname citenamefont\endcsname\relax
  \def\citenamefont#1{#1}\fi
\expandafter\ifx\csname url\endcsname\relax
  \def\url#1{\texttt{#1}}\fi
\expandafter\ifx\csname urlprefix\endcsname\relax\def\urlprefix{URL }\fi
\providecommand{\bibinfo}[2]{#2}
\providecommand{\eprint}[2][]{\url{#2}}

\bibitem[{\citenamefont{Udem et~al.}(2002)\citenamefont{Udem, Holzwarth, and
  H\"ansch}}]{udem_optical_2002}
\bibinfo{author}{\bibfnamefont{T.}~\bibnamefont{Udem}},
  \bibinfo{author}{\bibfnamefont{R.}~\bibnamefont{Holzwarth}},
  \bibnamefont{and} \bibinfo{author}{\bibfnamefont{T.~W.}
  \bibnamefont{H\"ansch}}, \bibinfo{journal}{Nature}
  \textbf{\bibinfo{volume}{416}}, \bibinfo{pages}{233} (\bibinfo{year}{2002}),
  ISSN \bibinfo{issn}{0028-0836}.

\bibitem[{\citenamefont{Diddams}(2010)}]{diddams2010evolving}
\bibinfo{author}{\bibfnamefont{S.~A.} \bibnamefont{Diddams}},
  \bibinfo{journal}{JOSA B} \textbf{\bibinfo{volume}{27}}, \bibinfo{pages}{B51}
  (\bibinfo{year}{2010}).

\bibitem[{\citenamefont{Schiller}(2002)}]{schiller2002spectrometry}
\bibinfo{author}{\bibfnamefont{S.}~\bibnamefont{Schiller}},
  \bibinfo{journal}{Optics letters} \textbf{\bibinfo{volume}{27}},
  \bibinfo{pages}{766} (\bibinfo{year}{2002}).

\bibitem[{\citenamefont{Keilmann et~al.}(2004)\citenamefont{Keilmann, Gohle,
  and Holzwarth}}]{keilmann2004time}
\bibinfo{author}{\bibfnamefont{F.}~\bibnamefont{Keilmann}},
  \bibinfo{author}{\bibfnamefont{C.}~\bibnamefont{Gohle}}, \bibnamefont{and}
  \bibinfo{author}{\bibfnamefont{R.}~\bibnamefont{Holzwarth}},
  \bibinfo{journal}{Optics Letters} \textbf{\bibinfo{volume}{29}},
  \bibinfo{pages}{1542} (\bibinfo{year}{2004}).

\bibitem[{\citenamefont{Coddington et~al.}(2008)\citenamefont{Coddington,
  Swann, and Newbury}}]{coddington2008coherent}
\bibinfo{author}{\bibfnamefont{I.}~\bibnamefont{Coddington}},
  \bibinfo{author}{\bibfnamefont{W.~C.} \bibnamefont{Swann}}, \bibnamefont{and}
  \bibinfo{author}{\bibfnamefont{N.~R.} \bibnamefont{Newbury}},
  \bibinfo{journal}{Physical review letters} \textbf{\bibinfo{volume}{100}},
  \bibinfo{pages}{013902} (\bibinfo{year}{2008}).

\bibitem[{\citenamefont{Coddington et~al.}(2010)\citenamefont{Coddington,
  Swann, and Newbury}}]{coddington2010coherent}
\bibinfo{author}{\bibfnamefont{I.}~\bibnamefont{Coddington}},
  \bibinfo{author}{\bibfnamefont{W.}~\bibnamefont{Swann}}, \bibnamefont{and}
  \bibinfo{author}{\bibfnamefont{N.}~\bibnamefont{Newbury}},
  \bibinfo{journal}{Physical Review A} \textbf{\bibinfo{volume}{82}},
  \bibinfo{pages}{043817} (\bibinfo{year}{2010}).

\bibitem[{\citenamefont{Bernhardt
  et~al.}(2010{\natexlab{a}})\citenamefont{Bernhardt, Ozawa, Jacquet, Jacquey,
  Kobayashi, Udem, Holzwarth, Guelachvili, H{\"a}nsch, and
  Picqu{\'e}}}]{bernhardt2009cavity}
\bibinfo{author}{\bibfnamefont{B.}~\bibnamefont{Bernhardt}},
  \bibinfo{author}{\bibfnamefont{A.}~\bibnamefont{Ozawa}},
  \bibinfo{author}{\bibfnamefont{P.}~\bibnamefont{Jacquet}},
  \bibinfo{author}{\bibfnamefont{M.}~\bibnamefont{Jacquey}},
  \bibinfo{author}{\bibfnamefont{Y.}~\bibnamefont{Kobayashi}},
  \bibinfo{author}{\bibfnamefont{T.}~\bibnamefont{Udem}},
  \bibinfo{author}{\bibfnamefont{R.}~\bibnamefont{Holzwarth}},
  \bibinfo{author}{\bibfnamefont{G.}~\bibnamefont{Guelachvili}},
  \bibinfo{author}{\bibfnamefont{T.~W.} \bibnamefont{H{\"a}nsch}},
  \bibnamefont{and}
  \bibinfo{author}{\bibfnamefont{N.}~\bibnamefont{Picqu{\'e}}},
  \bibinfo{journal}{Nature photonics} \textbf{\bibinfo{volume}{4}},
  \bibinfo{pages}{55} (\bibinfo{year}{2010}{\natexlab{a}}).

\bibitem[{\citenamefont{Schliesser et~al.}(2012)\citenamefont{Schliesser,
  Picqu{\'e}, and H{\"a}nsch}}]{schliesser2012mid}
\bibinfo{author}{\bibfnamefont{A.}~\bibnamefont{Schliesser}},
  \bibinfo{author}{\bibfnamefont{N.}~\bibnamefont{Picqu{\'e}}},
  \bibnamefont{and} \bibinfo{author}{\bibfnamefont{T.~W.}
  \bibnamefont{H{\"a}nsch}}, \bibinfo{journal}{Nature Photonics}
  \textbf{\bibinfo{volume}{6}}, \bibinfo{pages}{440} (\bibinfo{year}{2012}).

\bibitem[{\citenamefont{Bernhardt
  et~al.}(2010{\natexlab{b}})\citenamefont{Bernhardt, Sorokin, Jacquet, Thon,
  Becker, Sorokina, Picqu{\'e}, and H{\"a}nsch}}]{bernhardt2010mid}
\bibinfo{author}{\bibfnamefont{B.}~\bibnamefont{Bernhardt}},
  \bibinfo{author}{\bibfnamefont{E.}~\bibnamefont{Sorokin}},
  \bibinfo{author}{\bibfnamefont{P.}~\bibnamefont{Jacquet}},
  \bibinfo{author}{\bibfnamefont{R.}~\bibnamefont{Thon}},
  \bibinfo{author}{\bibfnamefont{T.}~\bibnamefont{Becker}},
  \bibinfo{author}{\bibfnamefont{I.}~\bibnamefont{Sorokina}},
  \bibinfo{author}{\bibfnamefont{N.}~\bibnamefont{Picqu{\'e}}},
  \bibnamefont{and}
  \bibinfo{author}{\bibfnamefont{T.}~\bibnamefont{H{\"a}nsch}},
  \bibinfo{journal}{Applied Physics B} \textbf{\bibinfo{volume}{100}},
  \bibinfo{pages}{3} (\bibinfo{year}{2010}{\natexlab{b}}).

\bibitem[{\citenamefont{Baumann et~al.}(2011)\citenamefont{Baumann, Giorgetta,
  Swann, Zolot, Coddington, and Newbury}}]{baumann2011spectroscopy}
\bibinfo{author}{\bibfnamefont{E.}~\bibnamefont{Baumann}},
  \bibinfo{author}{\bibfnamefont{F.}~\bibnamefont{Giorgetta}},
  \bibinfo{author}{\bibfnamefont{W.}~\bibnamefont{Swann}},
  \bibinfo{author}{\bibfnamefont{A.}~\bibnamefont{Zolot}},
  \bibinfo{author}{\bibfnamefont{I.}~\bibnamefont{Coddington}},
  \bibnamefont{and} \bibinfo{author}{\bibfnamefont{N.}~\bibnamefont{Newbury}},
  \bibinfo{journal}{Physical Review A} \textbf{\bibinfo{volume}{84}},
  \bibinfo{pages}{062513} (\bibinfo{year}{2011}).

\bibitem[{\citenamefont{Giorgetta et~al.}(2015)\citenamefont{Giorgetta, Rieker,
  Baumann, Swann, Sinclair, Kofler, Coddington, and
  Newbury}}]{giorgetta2015broadband}
\bibinfo{author}{\bibfnamefont{F.~R.} \bibnamefont{Giorgetta}},
  \bibinfo{author}{\bibfnamefont{G.~B.} \bibnamefont{Rieker}},
  \bibinfo{author}{\bibfnamefont{E.}~\bibnamefont{Baumann}},
  \bibinfo{author}{\bibfnamefont{W.~C.} \bibnamefont{Swann}},
  \bibinfo{author}{\bibfnamefont{L.~C.} \bibnamefont{Sinclair}},
  \bibinfo{author}{\bibfnamefont{J.}~\bibnamefont{Kofler}},
  \bibinfo{author}{\bibfnamefont{I.}~\bibnamefont{Coddington}},
  \bibnamefont{and} \bibinfo{author}{\bibfnamefont{N.~R.}
  \bibnamefont{Newbury}}, \bibinfo{journal}{Physical review letters}
  \textbf{\bibinfo{volume}{115}}, \bibinfo{pages}{103901}
  (\bibinfo{year}{2015}).

\bibitem[{\citenamefont{Adler et~al.}(2009)\citenamefont{Adler, Cossel, Thorpe,
  Hartl, Fermann, and Ye}}]{Adler:2009}
\bibinfo{author}{\bibfnamefont{F.}~\bibnamefont{Adler}},
  \bibinfo{author}{\bibfnamefont{K.~C.} \bibnamefont{Cossel}},
  \bibinfo{author}{\bibfnamefont{M.~J.} \bibnamefont{Thorpe}},
  \bibinfo{author}{\bibfnamefont{I.}~\bibnamefont{Hartl}},
  \bibinfo{author}{\bibfnamefont{M.~E.} \bibnamefont{Fermann}},
  \bibnamefont{and} \bibinfo{author}{\bibfnamefont{J.}~\bibnamefont{Ye}},
  \bibinfo{journal}{Opt. Lett.} \textbf{\bibinfo{volume}{34}},
  \bibinfo{pages}{1330} (\bibinfo{year}{2009}).

\bibitem[{\citenamefont{Vodopyanov et~al.}(2011)\citenamefont{Vodopyanov,
  Sorokin, Sorokina, and Schunemann}}]{Vodopyanov:2011}
\bibinfo{author}{\bibfnamefont{K.~L.} \bibnamefont{Vodopyanov}},
  \bibinfo{author}{\bibfnamefont{E.}~\bibnamefont{Sorokin}},
  \bibinfo{author}{\bibfnamefont{I.~T.} \bibnamefont{Sorokina}},
  \bibnamefont{and} \bibinfo{author}{\bibfnamefont{P.~G.}
  \bibnamefont{Schunemann}}, \bibinfo{journal}{Opt. Lett.}
  \textbf{\bibinfo{volume}{36}}, \bibinfo{pages}{2275} (\bibinfo{year}{2011}).

\bibitem[{\citenamefont{Galli et~al.}(2013)\citenamefont{Galli, Cappelli,
  Cancio, Giusfredi, Mazzotti, Bartalini, and De~Natale}}]{Galli:2013d}
\bibinfo{author}{\bibfnamefont{I.}~\bibnamefont{Galli}},
  \bibinfo{author}{\bibfnamefont{F.}~\bibnamefont{Cappelli}},
  \bibinfo{author}{\bibfnamefont{P.}~\bibnamefont{Cancio}},
  \bibinfo{author}{\bibfnamefont{G.}~\bibnamefont{Giusfredi}},
  \bibinfo{author}{\bibfnamefont{D.}~\bibnamefont{Mazzotti}},
  \bibinfo{author}{\bibfnamefont{S.}~\bibnamefont{Bartalini}},
  \bibnamefont{and}
  \bibinfo{author}{\bibfnamefont{P.}~\bibnamefont{De~Natale}},
  \bibinfo{journal}{Opt. Express} \textbf{\bibinfo{volume}{21}},
  \bibinfo{pages}{28877} (\bibinfo{year}{2013}).

\bibitem[{\citenamefont{Cruz et~al.}(2015)\citenamefont{Cruz, Maser, Johnson,
  Ycas, Klose, Giorgetta, Coddington, and Diddams}}]{cruz2015mid}
\bibinfo{author}{\bibfnamefont{F.~C.} \bibnamefont{Cruz}},
  \bibinfo{author}{\bibfnamefont{D.~L.} \bibnamefont{Maser}},
  \bibinfo{author}{\bibfnamefont{T.}~\bibnamefont{Johnson}},
  \bibinfo{author}{\bibfnamefont{G.}~\bibnamefont{Ycas}},
  \bibinfo{author}{\bibfnamefont{A.}~\bibnamefont{Klose}},
  \bibinfo{author}{\bibfnamefont{F.~R.} \bibnamefont{Giorgetta}},
  \bibinfo{author}{\bibfnamefont{I.}~\bibnamefont{Coddington}},
  \bibnamefont{and} \bibinfo{author}{\bibfnamefont{S.~A.}
  \bibnamefont{Diddams}}, \bibinfo{journal}{Optics Express}
  \textbf{\bibinfo{volume}{23}}, \bibinfo{pages}{26814} (\bibinfo{year}{2015}).

\bibitem[{\citenamefont{Faist et~al.}(1994)\citenamefont{Faist, Capasso, Sivco,
  Sirtori, Hutchinson, and Cho}}]{faist1994quantum}
\bibinfo{author}{\bibfnamefont{J.}~\bibnamefont{Faist}},
  \bibinfo{author}{\bibfnamefont{F.}~\bibnamefont{Capasso}},
  \bibinfo{author}{\bibfnamefont{D.~L.} \bibnamefont{Sivco}},
  \bibinfo{author}{\bibfnamefont{C.}~\bibnamefont{Sirtori}},
  \bibinfo{author}{\bibfnamefont{A.~L.} \bibnamefont{Hutchinson}},
  \bibnamefont{and} \bibinfo{author}{\bibfnamefont{A.~Y.} \bibnamefont{Cho}},
  \bibinfo{journal}{Science} \textbf{\bibinfo{volume}{264}},
  \bibinfo{pages}{553} (\bibinfo{year}{1994}).

\bibitem[{\citenamefont{Hugi et~al.}(2012)\citenamefont{Hugi, Villares, Blaser,
  Liu, and Faist}}]{hugi2012mid}
\bibinfo{author}{\bibfnamefont{A.}~\bibnamefont{Hugi}},
  \bibinfo{author}{\bibfnamefont{G.}~\bibnamefont{Villares}},
  \bibinfo{author}{\bibfnamefont{S.}~\bibnamefont{Blaser}},
  \bibinfo{author}{\bibfnamefont{H.}~\bibnamefont{Liu}}, \bibnamefont{and}
  \bibinfo{author}{\bibfnamefont{J.}~\bibnamefont{Faist}},
  \bibinfo{journal}{Nature} \textbf{\bibinfo{volume}{492}},
  \bibinfo{pages}{229} (\bibinfo{year}{2012}).

\bibitem[{\citenamefont{Burghoff et~al.}(2014)\citenamefont{Burghoff, Kao, Han,
  Chan, Cai, Yang, Hayton, Gao, Reno, and Hu}}]{Burghoff:2014}
\bibinfo{author}{\bibfnamefont{D.}~\bibnamefont{Burghoff}},
  \bibinfo{author}{\bibfnamefont{T.-Y.} \bibnamefont{Kao}},
  \bibinfo{author}{\bibfnamefont{N.}~\bibnamefont{Han}},
  \bibinfo{author}{\bibfnamefont{C.~W.~I.} \bibnamefont{Chan}},
  \bibinfo{author}{\bibfnamefont{X.}~\bibnamefont{Cai}},
  \bibinfo{author}{\bibfnamefont{Y.}~\bibnamefont{Yang}},
  \bibinfo{author}{\bibfnamefont{D.~J.} \bibnamefont{Hayton}},
  \bibinfo{author}{\bibfnamefont{J.-R.} \bibnamefont{Gao}},
  \bibinfo{author}{\bibfnamefont{J.~L.} \bibnamefont{Reno}}, \bibnamefont{and}
  \bibinfo{author}{\bibfnamefont{Q.}~\bibnamefont{Hu}}, \bibinfo{journal}{Nat.
  Photon.} \textbf{\bibinfo{volume}{8}}, \bibinfo{pages}{462}
  (\bibinfo{year}{2014}).

\bibitem[{\citenamefont{R{\"o}sch et~al.}(2015)\citenamefont{R{\"o}sch,
  Scalari, Beck, and Faist}}]{rosch2015octave}
\bibinfo{author}{\bibfnamefont{M.}~\bibnamefont{R{\"o}sch}},
  \bibinfo{author}{\bibfnamefont{G.}~\bibnamefont{Scalari}},
  \bibinfo{author}{\bibfnamefont{M.}~\bibnamefont{Beck}}, \bibnamefont{and}
  \bibinfo{author}{\bibfnamefont{J.}~\bibnamefont{Faist}},
  \bibinfo{journal}{Nature Photonics} \textbf{\bibinfo{volume}{9}},
  \bibinfo{pages}{42} (\bibinfo{year}{2015}).

\bibitem[{\citenamefont{Lu et~al.}(2015)\citenamefont{Lu, Razeghi, Slivken,
  Bandyopadhyay, Bai, Zhou, Chen, Heydari, Haddadi, McClintock
  et~al.}}]{lu2015high}
\bibinfo{author}{\bibfnamefont{Q.}~\bibnamefont{Lu}},
  \bibinfo{author}{\bibfnamefont{M.}~\bibnamefont{Razeghi}},
  \bibinfo{author}{\bibfnamefont{S.}~\bibnamefont{Slivken}},
  \bibinfo{author}{\bibfnamefont{N.}~\bibnamefont{Bandyopadhyay}},
  \bibinfo{author}{\bibfnamefont{Y.}~\bibnamefont{Bai}},
  \bibinfo{author}{\bibfnamefont{W.}~\bibnamefont{Zhou}},
  \bibinfo{author}{\bibfnamefont{M.}~\bibnamefont{Chen}},
  \bibinfo{author}{\bibfnamefont{D.}~\bibnamefont{Heydari}},
  \bibinfo{author}{\bibfnamefont{A.}~\bibnamefont{Haddadi}},
  \bibinfo{author}{\bibfnamefont{R.}~\bibnamefont{McClintock}},
  \bibnamefont{et~al.}, \bibinfo{journal}{Applied Physics Letters}
  \textbf{\bibinfo{volume}{106}}, \bibinfo{pages}{051105}
  (\bibinfo{year}{2015}).

\bibitem[{\citenamefont{Khurgin et~al.}(2014)\citenamefont{Khurgin, Dikmelik,
  Hugi, and Faist}}]{Khurgin:2014}
\bibinfo{author}{\bibfnamefont{J.~B.} \bibnamefont{Khurgin}},
  \bibinfo{author}{\bibfnamefont{Y.}~\bibnamefont{Dikmelik}},
  \bibinfo{author}{\bibfnamefont{A.}~\bibnamefont{Hugi}}, \bibnamefont{and}
  \bibinfo{author}{\bibfnamefont{J.}~\bibnamefont{Faist}},
  \bibinfo{journal}{Appl. Phys. Lett.} \textbf{\bibinfo{volume}{104}},
  \bibinfo{pages}{081118} (\bibinfo{year}{2014}).

\bibitem[{\citenamefont{Villares and Faist}(2015)}]{Villares:2015}
\bibinfo{author}{\bibfnamefont{G.}~\bibnamefont{Villares}} \bibnamefont{and}
  \bibinfo{author}{\bibfnamefont{J.}~\bibnamefont{Faist}},
  \bibinfo{journal}{Opt. Express} \textbf{\bibinfo{volume}{23}},
  \bibinfo{pages}{1651} (\bibinfo{year}{2015}).

\bibitem[{\citenamefont{Villares et~al.}(2015)\citenamefont{Villares, Riedi,
  Wolf, Kazakov, S{\"u}ess, Beck, and Faist}}]{villares2015dispersion}
\bibinfo{author}{\bibfnamefont{G.}~\bibnamefont{Villares}},
  \bibinfo{author}{\bibfnamefont{S.}~\bibnamefont{Riedi}},
  \bibinfo{author}{\bibfnamefont{J.}~\bibnamefont{Wolf}},
  \bibinfo{author}{\bibfnamefont{D.}~\bibnamefont{Kazakov}},
  \bibinfo{author}{\bibfnamefont{M.~J.} \bibnamefont{S{\"u}ess}},
  \bibinfo{author}{\bibfnamefont{M.}~\bibnamefont{Beck}}, \bibnamefont{and}
  \bibinfo{author}{\bibfnamefont{J.}~\bibnamefont{Faist}},
  \bibinfo{journal}{arXiv preprint arXiv:1509.08856}  (\bibinfo{year}{2015}).

\bibitem[{\citenamefont{Villares et~al.}(2014)\citenamefont{Villares, Hugi,
  Blaser, and Faist}}]{Villares:2014}
\bibinfo{author}{\bibfnamefont{G.}~\bibnamefont{Villares}},
  \bibinfo{author}{\bibfnamefont{A.}~\bibnamefont{Hugi}},
  \bibinfo{author}{\bibfnamefont{S.}~\bibnamefont{Blaser}}, \bibnamefont{and}
  \bibinfo{author}{\bibfnamefont{J.}~\bibnamefont{Faist}},
  \bibinfo{journal}{Nat. Commun.} \textbf{\bibinfo{volume}{5}},
  \bibinfo{pages}{5192} (\bibinfo{year}{2014}).

\bibitem[{\citenamefont{Bismuto et~al.}(2015)\citenamefont{Bismuto, Bidaux,
  Tardy, Terazzi, Gresch, Wolf, Blaser, M{\"u}ller, and
  Faist}}]{bismuto2015electricaltuning}
\bibinfo{author}{\bibfnamefont{A.}~\bibnamefont{Bismuto}},
  \bibinfo{author}{\bibfnamefont{Y.}~\bibnamefont{Bidaux}},
  \bibinfo{author}{\bibfnamefont{C.}~\bibnamefont{Tardy}},
  \bibinfo{author}{\bibfnamefont{R.}~\bibnamefont{Terazzi}},
  \bibinfo{author}{\bibfnamefont{T.}~\bibnamefont{Gresch}},
  \bibinfo{author}{\bibfnamefont{J.}~\bibnamefont{Wolf}},
  \bibinfo{author}{\bibfnamefont{S.}~\bibnamefont{Blaser}},
  \bibinfo{author}{\bibfnamefont{A.}~\bibnamefont{M{\"u}ller}},
  \bibnamefont{and} \bibinfo{author}{\bibfnamefont{J.}~\bibnamefont{Faist}},
  \bibinfo{journal}{Optics Express, in press}  (\bibinfo{year}{2015}).

\bibitem[{\citenamefont{Sergachev et~al.}(2014)\citenamefont{Sergachev,
  Maulini, Bismuto, Blaser, Gresch, Bidaux, M{\"u}ller, Schilt, and
  S{\"u}dmeyer}}]{sergachev2014all}
\bibinfo{author}{\bibfnamefont{I.}~\bibnamefont{Sergachev}},
  \bibinfo{author}{\bibfnamefont{R.}~\bibnamefont{Maulini}},
  \bibinfo{author}{\bibfnamefont{A.}~\bibnamefont{Bismuto}},
  \bibinfo{author}{\bibfnamefont{S.}~\bibnamefont{Blaser}},
  \bibinfo{author}{\bibfnamefont{T.}~\bibnamefont{Gresch}},
  \bibinfo{author}{\bibfnamefont{Y.}~\bibnamefont{Bidaux}},
  \bibinfo{author}{\bibfnamefont{A.}~\bibnamefont{M{\"u}ller}},
  \bibinfo{author}{\bibfnamefont{S.}~\bibnamefont{Schilt}}, \bibnamefont{and}
  \bibinfo{author}{\bibfnamefont{T.}~\bibnamefont{S{\"u}dmeyer}},
  \bibinfo{journal}{Optics letters} \textbf{\bibinfo{volume}{39}},
  \bibinfo{pages}{6411} (\bibinfo{year}{2014}).

\end{thebibliography}

\end{document}